\documentclass[twocolumn,showpacs,preprintnumbers,amsmath,amssymb]{revtex4}
\usepackage{graphicx}% Include figure files
\usepackage{bm}% bold math

\newcommand{\ie}{{\it i.e.}}

\def\ket#1{\mid\!#1\rangle}

\begin{document}

\title{Conversion of $^{40}$K-$^{87}$Rb mixtures into stable molecules}

\author{Li-Hua Lu and You-Quan Li}
\affiliation{Zhejiang Institute of Modern Physics and Department of Physics,\\
Zhejiang University, Hangzhou 310027, P. R. China}

\begin{abstract}
We study the conversion of $^{40}$K and $^{87}$Rb atoms into
stable molecules through
the stimulated Raman adiabatic passage (STIRAP) in photoassociation
assisted with Feshbach resonance.
Starting with the mean-field Langrange density, we show that the
atom-to-molecule conversion efficiency by STIRAP aided by Feshbach
resonance is much larger than that by bare Feshbach resonance. We
also study the influence of the population imbalance on the
atom-to-molecule conversion.
\end{abstract}
\received{19 May 2007; revised manuscript received 22 August 2007}
%\received{\today}
\pacs{03.75.Lm, 03.75.Hh}
%03.75.Lm Tunneling, Josephson effect, Bose-Einstein condensates in periodic potentials, solitons, vortices and topological excitations
%03.75.Hh Static properties of condensates; thermodynamical, statistical and structural properties
%03.75.Kk Dynamic properties of condensates; collective and hydrodynamic excitations, superfluid flow

\maketitle

\section{Introduction}

The study on cold atoms is a remarkable research area which has
been extended from monoatomic systems to diatomic systems in recent
years~\cite{Dema,OHara, Schr, Hadz, Modu, Roati, Strecker, Ospel}.
In these experiments, not only  degenerate
Fermi-Fermi~\cite{Dema,OHara,Strecker} but also
Fermi-Bose~\cite{Schr, Hadz, Modu, Roati,Ospel} mixtures are
studied.
Principally, the degenerate atoms can be converted into molecules
through the resonant photoassociation or magnetoassociation
(Fechbash resonance). Note that such a compound
molecule created by the Fechbash resonance is in a quasibound
state and hence energetically unstable.
Although the quasibound
molecules are energetically unstable, most of fermionic atoms can
be efficiently converted into molecules with a long lifetime in
experiments~\cite{Greiner,Zwier, Regal}.
However, as far as we
know, the bare Feshbach resonance can not convert either
bosonic atoms or Fermi-Bose mixtures into molecules with high efficiency.

To investigate various novel features of cold molecules, one must
create ground-state molecules with high atom-to-molecule
conversion efficiency. The bare stimulated Raman adiabatic passage
(STIRAP)~\cite{Vardi,Bergmann} in photoassociation is proposed to
enhance the atom-to-molecule conversion efficiency through
avoiding quasibound molecules' radiation decay. The success of
the STIRAP technique relies on the existence of the coherent
population trapping (CPT) state~\cite{Alze}, \ie, the system can
evolve adiabatically in coherent superposition of stable states.
It is easily satisfied for non-interaction systems.
However,
the inter-particle collisions make it difficult
for interaction systems to evolve adiabatically,
which limits a practical efficiency of STIRAP.
Comparing  with the bare STIRAP in
photoassociation, the stimulated Raman adiabatic passage aided by
magnetoassociation is found to be a more efficient technique whose
conversion efficiency is not limited by the collisions between
atoms~\cite{M. F.,Mackie,Ling}.
This technique can be applied to a
monoatomic system or a diatomic system.
The free atomic states together with the quasibound and ground molecular
states constitute a four-level system
for which STIRAP aided by magnetoassociation is applicable.

In this paper, we consider systems consisting of fermionic and
bosonic atoms and their compounded fermionic molecules through the
technique of STIRAP aided by magnetoassociation. We take the
$^{40}$K-$^{87}$Rb mixture as an example to show that this technique
can convert the two species of atoms into molecules with high
efficiency. In comparison with STIRAP aided by magnetoassociation,
we also show that the bare Feshbach resonance can hardly convert
$^{40}$K-$^{87}$Rb  atoms into molecules.  The influence of the
population imbalance between two species on the atom-to-molecule
conversion efficiency is also studied. In the next section, we
present our model. In Sec. \ref{sec:CPT} we derive the mean-field
dynamical equations through the Euler-Lagrange equation then
obtain the CPT solutions and the corresponding "two-photon"
resonance condition. In Sec. \ref{sec:num} we solve the dynamical
equations numerically and discuss the corresponding results. Our
main conclusions are summarized in Sec. \ref{sec:con}.

\section{Theoretical model}\label{sec:model}

We consider a  mixture of fermionic and bosonic atoms which are
coupled to a quasibound molecular state via Feshbach resonance.
Meanwhile, a laser field drives transitions between the quasibound
and ground molecular states. For convenience, let $\ket{f}$ and
$\ket{b}$ stand for the ground states of fermionic and bosonic
atoms in the open channel, and $\ket{m}$ and $\ket{g}$ for the
quasibound and ground molecular states in the close channel,
respectively. The state $\ket{m}$ is coupled  with states
$\ket{f}$ and $\ket{b}$ through a magnetic field with coupling
strength $\alpha'$ and detuning $\mathcal{E}'$. Additionally, the
states $\ket{m}$ and $\ket{g}$ are coupled with each other through a
laser field with the coupling strength $\Omega'$ and detuning
$\Delta'$. Then the Hamiltonian describing the above system in the
interaction picture is written as,
\begin{eqnarray}\label{eq:Hamiltonian}
\displaystyle\hat{H}&=&\int d\mathbf{r}
 \Bigl\{\sum_i T_i\hat{\Psi}^+_{i}\hat{\Psi}^{}_{i}
 +\frac{1}{2}\sum_{i,j}\lambda'_{ij}\hat{\Psi}^+_{i}
 \hat{\Psi}^+_{j}\hat{\Psi}^{}_{j}\hat{\Psi}^{}_{i}
   \nonumber\\
&& +\frac{\alpha'}{2}[\hat{\Psi}^+_m \hat{\Psi}^{}_f \hat{\Psi}^{}_b + H.c.]
+(\Delta'+\mathcal{E}')\hat{\Psi}^+_g\hat{\Psi}^{}_g
     \nonumber\\
&& +\mathcal{E}'\hat{\Psi}^+_m\hat{\Psi}^{}_m
-\frac{\Omega'}{2}[\hat{\Psi}^+_m\hat{\Psi}^{}_g + H.c.]\Bigr\},
\end{eqnarray}
where $\hat{\Psi}^{}_{i(j)}$ and $\hat{\Psi}_{i(j)}^+$ ($i,\, j
=f, b, m$ and $g$) are the annihilation and creation field
operators. They obey the commutation (+) or anticommutation (-)
relations
$[\hat{\Psi}_i(\mathbf{r},t),\hat{\Psi}_j^+(\mathbf{r},t)]_{\pm}
=\delta_{ij}\delta(\mathbf{r}-\mathbf{r}')$ for bosons or
fermions, respectively.  The coefficient $T_i$ is the kinetic
energy due to the particles' motions, and
$\lambda_{ij}'=\lambda_{ji}'=2\pi\hbar^2a^{}_{ij}/m^{}_{ij}$  the
interaction strength between particles with $a^{}_{ij}$ being the
$s$-wave scattering length and
$m^{}_{ij}=m^{}_im^{}_j/(m^{}_i+m^{}_j)$ being the reduced mass.
The trapping potential term is not included in the Hamiltonian as
we merely consider a uniform system.

The system we considered includes
both bosonic and  fermionic components.
The self-interaction of the bosonic
component is distinctly  different from that of the fermionic one.
The kinetic energy dominates the intra-species interaction
for the fermionic component
as there is no $s$-wave scattering for two fermions in the same
internal state.
This is in
marked contrast to the bosonic component for which the
interaction energy dominates the kinetic one under most
experimental conditions.
Thus a very good first approximation is
neglecting the intra-species interaction for fermions but
neglecting the kinetic energy for bosons.
Based on the above consideration,
the energy density corresponding to the Hamiltonian
Eq.~(\ref{eq:Hamiltonian}), in the Hartree approximation, is given  by
\begin{eqnarray}
E&=&\displaystyle \frac{1}{2}\sum_{i\neq j}
\lambda'_{ij}|\psi_i|^2|\psi_j|^2 +\mathcal{E}'\psi_m^*\psi^{}_m
+(\Delta'+\mathcal{E}')\psi_g^*\psi^{}_g
  \nonumber\\
&&+\frac{\alpha'}{2}[\psi_m^*\psi_f\psi_b + H.c.]-\frac{\Omega'}{2}[\psi_m^*\psi_g+H.c.]\nonumber\\
&&+\frac{1}{2}\lambda_{bb}'|\psi_b|^4+\sum_{i=\{f,m,g\}}\frac{3}{5}A_i'|\psi_i|^{10/3},\nonumber\\
\end{eqnarray}
where $\psi_i$ represents the complex probability amplitude of the
$i$th component and $A_i'=\hbar^2(6\pi^2)^{2/3}/2m_i$. The
effective self-interaction  term $3A_i'|\psi_i|^{10/3}/5$ related
to fermions is called Pauli blocking term. One can find that  the
effective self-interaction is in different power of $|\psi|$ for
bosonic and fermionic components. This will induce distinct
difference between the dynamical equation for the bosonic
component and that for the fermionic one.

\section{Mean-field dynamics and CPT state}\label{sec:CPT}

The mean-field approach is an effective method to solve
many-body problems (particularly valid for systems with
a large number of particles)
although the high order quantum correlations are
ignored in this approximation~\cite{Parkins}.
There are different
approaches to get the mean-field dynamical equations from
Hamiltonian (\ref{eq:Hamiltonian}). A rigorous approach is
expanding bosonic and fermionic field operators in terms of
conventional creation and annihilation operators,
subsequently, substituting
them into the Heisenberg equations of motion for the field
operators (see Ref.~\cite{Salerno}).
In the mean-field
approximation, the evolution of each kind of bosons is determined
by a single equation, however, the evolution of $N_f$ fermions is
determined by $N_f$ equations.
Obviously, it is easy to handle the system with
bosons but difficult to handle the system containing a large
number of fermions. Another approach is based on the
Euler-Lagrange equation with the help of the mean-field Lagrangian
density~\cite{Capuzzi,Adhikari}. In this case, the approximation
attributes to a single evolution equation for each kind of
particles.
The difference between the mean-field dynamical equations
for bosons and that for fermions arises from the effective self-interaction term.

The method adopted in Refs.\cite{Ling,Salerno}
is not applicable to our system
due to the existence of fermions.
We study the
system in terms of mean-field Lagrange density,
\begin{equation}\label{eq:Lag}
\mathcal{L}=\displaystyle\frac{i}{2}\hbar\sum_i\Bigl(\psi^*_i\frac{\partial\psi_i}{
\partial t}-\psi_i\frac{\partial\psi_i^*}{\partial t}\Bigr)-E.
\end{equation}
Substituting the above mean-field Lagrangian density into the
Euler-Lagrange equation $\displaystyle\frac{\partial
\mathcal{L}}{\partial\psi_i^*}-\partial_\nu\Bigl(\frac{\partial\mathcal{L}}{\partial(\partial_\nu\psi_i^*)}
\Bigr)=0$, one can get a set of equations for the complex
probability amplitudes, $\psi^{}_f$, $\psi^{}_b$, $\psi^{}_m$,
and $\psi^{}_g$.
These equations are shown to guarantee the following
identities,
\begin{eqnarray*}
\frac{d}{dt}(|\psi^{}_f|^2+|\psi_m|^2+|\psi_g|^2)    &=& 0,\\
\frac{d}{dt}(|\psi^{}_b|^2+|\psi_m|^2+|\psi_g|^2) &=& 0,
\end{eqnarray*}
which means that the total numbers of species $f$ and $b$ are
conserved, \ie, $|\psi^{}_f|^2+|\psi_m|^2+|\psi_g|^2=n^{}_f$ and
$|\psi^{}_b|^2+|\psi_m|^2+|\psi_g|^2=n^{}_b$ with $n^{}_f$ and
$n^{}_b$ being constants. In the following discussion, we assume
there is no molecules in the system at the initial time, hence
$n^{}_f$ and $n^{}_b$ also denote the initial atom densities of
the corresponding species. To simplify the calculation, we let
$\phi_i=\psi_i/\sqrt{n^{}_f+n^{}_b}$, then the aforementioned
dynamical equations for $\psi_i$ become
\begin{eqnarray}\label{eq:dyna2}
i\frac{\partial \phi_f}{\partial t}&=&\sum_{i\neq
f}\lambda_{fi}|\phi^{}_i|^2\phi^{}_f+A_f(\phi_f^*\phi^{}_f)^{2/3}\phi^{}_f
+\frac{\alpha}{2}\phi_b^*\phi_m,\nonumber\\
i\frac{\partial \phi^{}_b}{\partial
t}&=&\sum_i\lambda_{bi}|\phi^{}_i|^2\phi^{}_b
+\frac{\alpha}{2}\phi_f^*\phi_m,\nonumber\\
i\frac{\partial\phi_m}{\partial t}&=&\sum_{i\neq
m}\lambda_{mi}|\phi_{i}^{}|^2\phi_m+A_m(\phi_m^*\phi_m)^{2/3}\phi_m+\frac{\alpha}{2}\phi^{}_f\phi^{}_b
\nonumber\\
&&-\frac{\Omega}{2}\phi^{}_g+\mathcal{E}\phi_m,\nonumber\\
i\frac{\partial\phi^{}_g}{\partial t}&=&\sum_{i\neq
g}\lambda_{gi}|\phi^{}_{i}|^2\phi^{}_g+A_g(\phi_g^*\phi^{}_g)^{2/3}\phi^{}_g
-\frac{\Omega}{2}\phi_m\nonumber\\
&&+(\Delta+\mathcal{E})\phi^{}_g,
\end{eqnarray}
with the conservation relations
 $|\phi^{}_f|^2+|\phi_m|^2+|\phi^{}_g|^2=(1+\delta)\big/2$ and
 $|\phi^{}_b|^2+|\phi_m|^2+|\phi^{}_g|^2=(1-\delta)\big/2$.
Here we introduce $\delta=(n^{}_f-n^{}_b)\big/(n^{}_f+n^{}_b)$ to
characterize the population imbalance between species $f$ and $b$,
together with further simplified notions:
$\lambda_{ij}=\lambda_{ij}'(n^{}_f+n^{}_b)/\hbar$,
$\alpha=\alpha'\sqrt{n^{}_f+n^{}_b}/\hbar$,
$A_i=A_i'(n^{}_f+n^{}_b)^{2/3}/\hbar$, $\Omega=\Omega'/\hbar$,
$\mathcal{E}=\mathcal{E}'/\hbar$, and $\Delta=\Delta'/\hbar$. The
population imbalance  between two-species atoms is an important
parameter affecting the feature of the system. For example, the
population imbalance between fermionic atoms for different spin
state~\cite{Zwi} can induce the superfluid to normal state phase
transition.

Unlike the monoatomic system~\cite{Ling}, the densities of the
fermionic and bosonic atoms may be different, which will affect
the conventional two-photon resonance condition. Additionally, the
nonlinear terms in our system will be changed because of the Pauli
exclusive principle for fermions.  We will see that the STIRAP technique
aided by Feshbach resonance can convert the two species of atoms into molecules with
high efficiency, strictly in contrast to the situation with
Feshbach resonance only. We assume that Eqs.~(\ref{eq:dyna2})
support a CPT steady state with $\phi_m=0$ (the validity of this
assumption is supported by our result obtained in the following).
We search steady state solutions of Eqs.~(\ref{eq:dyna2}) with the
help of the following trial wave functions,
\begin{eqnarray}\label{eq:solu}
\displaystyle\phi^{}_{f,b}
 &=&|\phi^{}_{f,b}|e^{i\theta^{}_{f,b}}e^{-i\mu^{}_{f,b}t},\nonumber\\
\phi^{}_{m,g}&=&|\phi^{}_{m,g}|e^{i\theta^{}_{m,g}}e^{-i(\mu^{}_f+\mu^{}_b)t},
\end{eqnarray}
where $\mu^{}_f$ and $\mu^{}_b$ are undetermined parameters.
Substituting Eqs.~(\ref{eq:solu})
into Eqs.~(\ref{eq:dyna2}) and taking $|\phi_m|=0$,
one can find a set of solutions,
%\begin{widetext}
\begin{eqnarray}\label{eq:cpt}
\displaystyle |\phi^0_f|^2&=&\frac{(\alpha^2\delta-\Omega^2)
+\sqrt{(\alpha^2\delta-\Omega^2)^2+2\alpha^2\Omega^2(1+\delta)}}{2\alpha^2},
  \nonumber\\[2mm]
|\phi_b^0|^2&=&|\phi^0_f|^2-\delta, \,
|\phi_g^0|^2=\frac{1+\delta}{2}-|\phi^0_f|^2, \,
\theta_g =\theta_f+\theta^{}_b,
  \nonumber\\[1mm]
\mu^{}_f &=& \lambda_{fb}^{}|\phi^0_b|^2
  +\lambda_{fg}|\phi^0_g|^2+A_f^{}|\phi_f^0|^{4/3},
   \nonumber\\
\mu^{}_b &=& \lambda_{fb}^{}|\phi^0_f|^2
  +\lambda_{bb}|\phi^0_b|^2
   +\lambda_{bg}|\phi^0_g|^2,
\end{eqnarray}
together with a restriction condition
\begin{widetext}
\begin{equation}\label{eq:con}
\Delta=-\mathcal{E}+(\lambda^{}_{fb}-\lambda^{}_{fg})|\phi^0_f|^2
+(\lambda^{}_{bb}+\lambda^{}_{fb}-\lambda^{}_{bg})|\phi^0_b|^2+
(\lambda^{}_{fg}+\lambda^{}_{bg})|\phi^0_g|^2+A_f^{}|\phi^0_f|^{4/3}
-A^{}_g|\phi^0_g|^{4/3}.
\end{equation}
\end{widetext}
This is a result valid for converting mixtures of
fermionic and bossonic atoms into fermionic molecules.
The parameters $\lambda_{ij}$ and $A_i$ can be
obtained directly for concrete systems.

\section{Numerical results for  concrete systems }\label{sec:num}

Now we are in the position to consider a concrete system
consisting of two-species atoms (saying $^{40}$K and $^{87}$Rb)
and their compounded fermionic molecules. In order to carry out
the numerical calculations, it is necessary to fix magnitudes of
the parameters that affect the atom-to-molecule conversion
efficiency. As we know,
$\mathcal{E}'\approx(2\mu_e+\mu^{}_N)(B-B_m)$ where $\mu_e$ and
$\mu^{}_N$ refer to the electron and nucleus magnetic moments, $B$
the magnetic field (the resonances occur at $B=B_m$)~\cite{Los}.
To evaluate the parameter $\alpha'$, let us recall the form of the
$s$-wave scattering length between species $f$ and $b$:
$a^{}_{fb}=a\bigl(1-\Delta_{\mathrm{Fes}}/(B-B_m)\bigr)$. Here $a$
is the scattering length far off resonance and
$\Delta_{\mathrm{Fes}}=\alpha'^2m^{}_{fb}\Big/\bigl(2\pi\hbar^2a(2\mu_e+\mu^{}_N)\bigr)$
the resonance width with $m^{}_{fb}=m^{}_fm^{}_b/(m^{}_f+m^{}_b)$
being the reduced mass of species $f$ and $b$. The scattering
length $a$ for the $^{40}$K-$^{87}$Rb mixture determined through
Feshbach spectroscopy is  about $-185 a_0$ with $a_0$ being the
Bohr radius~\cite{Fer}, and the expectation value for the width
$\Delta_{\mathrm{Fes}}$ is about $-3$G~\cite{Zaccanti}. Thus we
obtain $\alpha'\approx9.07\times10^{-39}$J.

\begin{figure}[h]
\vspace{-7mm}
\includegraphics[width=81mm]{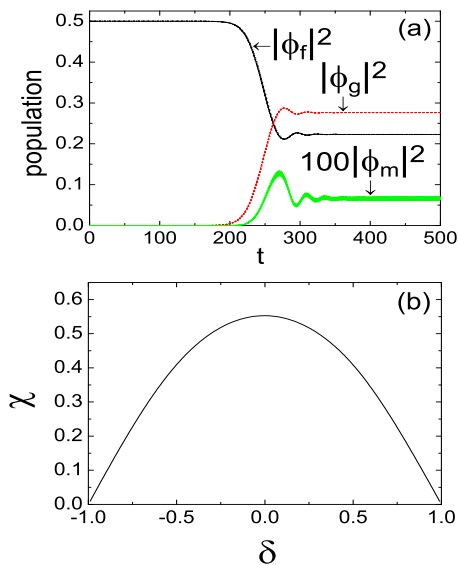}
\vspace{-10mm}
 \caption{\label{fig:bf} (color online) Panel (a)
shows the time dependence of the population of particles for
$\delta=0$. Panel (b) is the conversion efficiency
$\chi=2|\phi_g(t=\infty)|^2$ as the function of the population
imbalance $\delta$. The parameters are $\lambda^{}_{fb}=0.23$,
$\lambda_{bb}^{}=0.0056$,
$\lambda_{fm}^{}=\lambda^{}_{fg}=\lambda_{bm}=\lambda_{bg}^{}=\lambda_{mg}=0$,
$A_f^{}=0.3$, $A_m=A_g^{}=0.09$ and $\mathcal{E}=-4.1$. Time is in
unit of $1/\alpha$ and all other coefficients  are in units of
$\alpha$.}
\end{figure}

To optimize the conversion efficiency in experiment, one can
change the detuning $\mathcal{E}'$ and the interaction between
species $f$ and $b$ by varying the magnetic field B.
Alternatively, one can also fix the magnetic field but vary
$\Omega'$ to improve the conversion efficiency. In our numerical
calculation, the magnetic field $B$ is fixed to simplify the
experimental procedure. For $B=546.4$G, the corresponding $s$-wave
scattering length between species $f$ and $b$ is about 2950$a_0$.
If the total atomic density $n^{}_f+n^{}_b$ is about
$10^{20}$m$^{-3}$ and the $s$-wave scattering length between
$^{87}$Rb atoms is about 100$a_0$, it is easy to obtain
$\lambda^{}_{fb}=0.23\alpha$, $\lambda^{}_{bb}=0.0056\alpha$,
$A_f=0.3\alpha$, $A_m=A_g=0.09\alpha$ and
$\mathcal{E}=-4.1\alpha$. Since there are no good estimations on
molecular scattering lengths, we take the interaction
strengths involving molecules to be zero. The full set of
Eqs.~(\ref{eq:dyna2}) are solved numerically by choosing a
time-dependent Rabi frequency adopted in Ref.~\cite{Ling}
\begin{equation}
\Omega(t)=\Omega_{\mathrm{max}}\Bigl[1-\tanh\Bigl(\frac{t-t_0}{\tau}\Bigr)\Bigr],
\end{equation}
and $\Delta$  given by Eq.~(\ref{eq:con}). Here
$\Omega_{\mathrm{max}}$, $t_0$ and $\tau$ are constants to be
determined by the laser coupling of the two molecular states. We
take $\Omega_{\mathrm{max}}=200\alpha$, $t_0=120/\alpha$,
$\tau=40/\alpha$ and assume that there exist no molecules in the
system at the initial time, \ie, $\phi_{m,g}=0$ at $t=0$. Such an
assumption can be realized in experiment through the following
procedure.  At the initial time, fix the magnetic field at the
value far off resonance and prepare  $^{40}$K and $^{87}$Rb atoms
in their ground states $\ket{f}=\ket{F=9/2, m_F^{}=-9/2}$ and
$\ket{b}=\ket{1, 1}$, respectively. Turn on the laser pulse and
fix its amplitude at $\Omega_{\mathrm{max}}$, and then let the
magnetic field  be suddenly close to the Feshbash resonance point.
Some numerical results on atom-to-molecule conversion with
$B=546.4$G are plotted in Fig.~\ref{fig:bf}. Figure
\ref{fig:bf}(a) exhibits that $|\phi_m|^2\sim 0$ at the initial
time, which implies the populations fulfil the CPT solutions while
a small deviation begins to appear at about $t=300/\alpha$. In
spite of the small deviation, the final conversion efficiency
$2|\phi_g(t=\infty)|^2$ is about $60\%$. Note that the Feshbach
resonance can not  convert $^{40}$K-$^{87}$Rb atoms into molecules
with such a high efficiency. We plot the conversion efficiency as
a function of $\delta$ in Fig.~\ref{fig:bf}(b). Clearly, there is
a significant influence of $\delta$ on the conversion efficiency.

\begin{figure}[t]
\includegraphics[width=68mm]{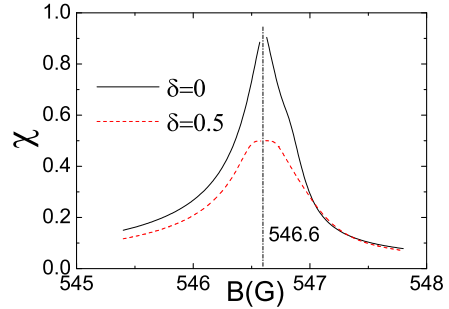}
\vspace{-3mm} \caption{\label{fig:proB} (color online) The
atom-to-molecule conversion efficiencies $2|\psi_g(t=\infty)|^2$
versus the magnetic field $B$ for different population imbalance
$\delta$. The parameters except  $\lambda_{fb}^{}$ and
$\mathcal{E}$ are the same as that in Fig. \ref{fig:bf} (a). }
\end{figure}

As we know, except $\lambda^{}_{fb}$ and $\mathcal{E}$,
the other parameters do not change with respect to the magnetic field.
Thus the magnetic field affects the conversion efficiency
through the detuning $\mathcal{E}$ and the interaction between species $f$ and $b$.
To show the influence of the magnetic field $B$,
we plot the atom-to-molecule conversion efficiency versus the magnetic
field in Fig. \ref{fig:proB}.
One can see that the strength of magnetic field can affect
the conversion efficiency distinctly and the curves are not
continuous at $B=546.6$G.
Such a discontinuity is brought in by the divergence of the interaction
$\lambda_{fb}^{}$ at $B=546.6$G.
\begin{figure}[t]
\includegraphics[width=70mm]{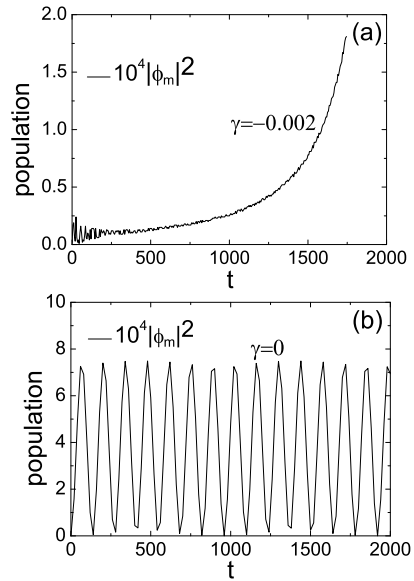}
\vspace{-3mm} \caption{\label{fig:Fesh} (color online) The time
dependence of molecular population for the Feshbach resonance. The
parameters are $\delta=0$, $\lambda^{}_{fb}=0.23$,
$\lambda_{bb}^{}=0.0056$, $\lambda_{fm}^{}=\lambda_{bm}=0$,
$A_f^{}=0.3$, $A_m=0.09$ and $\mathcal{E}=-4.1$. Time is in unit
of $1/\alpha$ and all other coefficients  are in units of
$\alpha$.}
\end{figure}

In comparison to the STIRAP aided by Feshbach resonance technique,
the results for the bare Feshbach resonance are plotted in Fig.
\ref{fig:Fesh}.
The results are obtained by solving the following set
of equations:
\begin{eqnarray}\label{eq:Feshbach}
i\frac{\partial \phi^{}_f}{\partial t}&=&\sum_{i=\{b,m\}
}\lambda_{fi}|\phi^{}_i|^2\phi^{}_f+A_f(\phi_f^*\phi^{}_f)^{2/3}\phi^{}_f
+\frac{\alpha}{2}\phi_b^*\phi_m,\nonumber\\
i\frac{\partial \phi^{}_b}{\partial
t}&=&\sum_{i=\{f,b,m\}}\lambda_{bi}|\phi^{}_i|^2\phi^{}_b
+\frac{\alpha}{2}\phi_f^*\phi_m,\nonumber\\
i\frac{\partial\phi_m}{\partial t}&=&\sum_{i=\{f,b\}
}\lambda_{mi}|\phi_{i}^{}|^2\phi_m+A_m(\phi_m^*\phi_m)^{2/3}\phi_m
\nonumber\\&&+\frac{\alpha}{2}\phi^{}_f\phi^{}_b
+\mathcal{E}\phi_m.
\end{eqnarray}
Unlike the STIRAP technique,
the magnetic field is swept according to $B=B_{\mathrm{ini}}+\gamma t$
(here $B_{\mathrm{ini}}$ denotes the initial value of the magnetic field)
in the Feshbach resonance approach.
The interaction $\lambda_{fb}$ and the detuning
$\mathcal{E}$ therefore change with time in the calculation procedure
if $\gamma \neq0$.
From Fig. \ref{fig:Fesh}, we find that the
Feshbach resonance technique can hardly convert $^{40}$K-$^{87}$Rb
atoms into molecules even if the magnetic field is swept slowly.
The magnetic field is brought from the initial value $551$G
to the final value $547.5$G in 2ms
in panel (a), whereas the magnetic field is fixed in panel (b)
due to $\gamma=0$.

\section{Summary}\label{sec:con}

We studied the atom-to-molecule conversion efficiency for
Fermi-Bose mixtures. We took the system consisting of
$^{40}$K-$^{87}$Rb atoms and their compounded molecules as an
example to have shown that the STIRAP aided by Feshbach resonance is an
efficient scheme for one to convert Fermi-Bose atoms into
fermionic molecules. Such a scheme is easy to handle in experiments
because it is no more necessary to sweep the magnetic field across
the Feshbach resonance. In contrast to the STIRAP aided by
Feshbach resonance technique, the bare Fershbach resonance can
hardly convert Fermi-Bose atoms into molecules. We showed that the
population imbalance $\delta$ between two-species atoms is a
conserved quantity which is an important parameter affecting the
feature of system.
We discussed the influence of $\delta$ on the
atom-to-molecule conversion in the system
that consists of Fermi-Bose atoms and their compounded molecules.

The work is supported by NSFC Grant No. 10674117.

\end{document}